\documentclass[prd,preprintnumbers,twocolumn,eqsecnum,floatfix,nofootinbib,a4paper,superscriptaddress]
{revtex4-1}
\usepackage{color}
\usepackage{calc}
\usepackage{amsmath,amssymb,graphicx}
\usepackage{amssymb,amsmath}
\usepackage{tensor}
\usepackage{bm}
\usepackage{float}
\usepackage{microtype}

\usepackage{booktabs}
\usepackage{times}
\usepackage[varg]{txfonts}
\usepackage[colorlinks, pdfborder={0 0 0}]{hyperref}
\usepackage{subfigure}
\definecolor{LinkColor}{rgb}{0.75, 0, 0}
\definecolor{CiteColor}{rgb}{0, 0.5, 0.5}
\definecolor{UrlColor}{rgb}{0, 0, 0.75}
\hypersetup{linkcolor=LinkColor}
\hypersetup{citecolor=CiteColor}
\hypersetup{urlcolor=UrlColor}
\allowdisplaybreaks
\usepackage[utf8]{inputenc}
\usepackage{ulem}
\DeclareFontFamily{OT1}{pzc}{}
\DeclareFontShape{OT1}{pzc}{m}{it}{<-> s * [1.10] pzcmi7t}{}
\DeclareMathAlphabet{\mathpzc}{OT1}{pzc}{m}{it}

\newcommand{\sk}[1]{}

\newcommand{\mnras}{Monthly Notices of the Royal Astronomical Society}

\newcommand{\Rmnum}[1]{\expandafter\@slowromancap\romannumeral #1@}
\makeatother

\def\mc{\mathcal}

\def\be{\begin{equation}}
\def\ee{\end{equation}}
\def\bea{\begin{eqnarray}}
\def\eea{\end{eqnarray}}
\def\bn{\begin{enumerate}}
\def\en{\end{enumerate}}
\def\bsube{\begin{subequations}}
\def\esube{\end{subequations}}

\def\H{{\mc{H}}}
\def\Hl{{\mc{H}_\textsc{l}}}
\def\Hu{{\mc{H}_\textsc{u}}}
\def\Ht{{\mc{H}_\textrm{t}}}
\def\Hs{{\mc{H}_\textrm{s}}}
\def\Hv{{\mc{H}_\textrm{v}}}
\def\btheta{{\bm \theta}}

\def\Ots{{\mc{O}_\textrm{s}^\textrm{t}}}

\def\Bts{{\mc{B}_\textrm{s}^\textrm{t}}}
\def\Btv{{\mc{B}_\textrm{v}^\textrm{t}}}

\def\Pts{{\mc{P}_\textrm{s}^\textrm{t}}}

\def\Blu{{\mc{B}_\textsc{u}^\textsc{l}}}

\def\e{\mathsf{e}}
\def\h{\mathsf{h}}
\def\D{\mathsf{D}}
\def\d{\mathbf{d}}

\begin{document}
	
\preprint{}
	
\title{Testing the nature of gravitational-wave polarizations using strongly lensed signals} 

\author{Srashti Goyal}
\affiliation{International Centre for Theoretical Sciences, Tata Institute of Fundamental Research, Bangalore 560089, India}
\author{K. Haris}
\affiliation{International Centre for Theoretical Sciences, Tata Institute of Fundamental Research, Bangalore 560089, India}
\affiliation{Nikhef – National Institute for Subatomic Physics, Science Park, 1098 XG Amsterdam, The Netherlands}
\author{Ajit Kumar Mehta}
\affiliation{International Centre for Theoretical Sciences, Tata Institute of Fundamental Research, Bangalore 560089, India}
\affiliation{Max Planck Institute for Gravitational Physics (Albert Einstein Institute), D-14476 Potsdam-Golm, Germany}
\author{Parameswaran~Ajith}
\affiliation{International Centre for Theoretical Sciences, Tata Institute of Fundamental Research, Bangalore 560089, India}
\affiliation{Canadian Institute for Advanced Research, CIFAR Azrieli Global Scholar, MaRS Centre, West Tower, 661 University Ave., Suite 505, Toronto, ON M5G 1M1, Canada}

\date{\today}
\begin{abstract}
Gravitational-wave (GW) observations by a network of ground-based laser interferometric detectors allow us to probe the nature of GW polarizations. This would be an interesting test of general relativity (GR), since GR predicts only two polarization modes while there are theories of gravity that predict up to six polarization modes. The ability of GW observations to probe the nature of polarizations is limited by the available number of linearly independent detectors in the network. (To extract all polarization modes, there should be at least as many detectors as the polarization modes.) Strong gravitational lensing of GWs offers a possibility to significantly increase the effective number of detectors in the network. Due to strong lensing (e.g., by galaxies), multiple copies of the same signal can be observed with time delays of several minutes to weeks. Owing to the rotation of the earth, observation of the multiple copies of the same GW signal would allow the network to measure different combinations of the same polarizations. This effectively multiplies the number of detectors in the network. Focusing on strongly lensed signals from binary black hole mergers that produce two observable ``images'', using Bayesian model selection and assuming simple polarization models, we show that our ability to distinguish between polarization models is significantly improved. 
\end{abstract}
	
	
	\maketitle
	
\section{Introduction}
Recent observations of gravitational waves (GW)~\cite{gw150914,gw170817,gwtc1,Venumadhav:2019lyq,Nitz:2019hdf} have offered new tests of general relativity (GR) in a regime inaccessible by other astronomical observations and laboratory tests~\cite{LSC_2016grtests,GWTC1TestofGR}. One set of interesting probes includes the nature of GWs themselves~\cite{Will:2014kxa}. For example, the near-simultaneous observations of the GW and gamma-ray signals from the binary neutron star merger GW170817~\cite{GW170817-GRB170817} have provided a stringent constraint on the speed of GWs, which, in turn, has ruled out several alternative theories to GR invoked to explain the accelerated expansion of the universe~\cite{Ezquiaga:2017ekz}; the amount of dispersion in the observed GW signals is constrained by bounding their deviation from GR-based templates, which in turn has provided an interesting upper bound on the mass of the graviton~\cite{Keppel:2010qu,LSC_2016grtests,GWTC1TestofGR}. Similarly, tests that constrain the values of various post-Newtonian parameters~\cite{Arun:2006hn,Mishra:2010tp,Li:2011cg,LSC_2016grtests,gw151226,GWTC1TestofGR} describing the GW signal also have constrained parameters of alternative theories~\cite{Yunes:2016jcc}.

\begin{figure}[tbh]
\includegraphics[width=0.95\columnwidth]{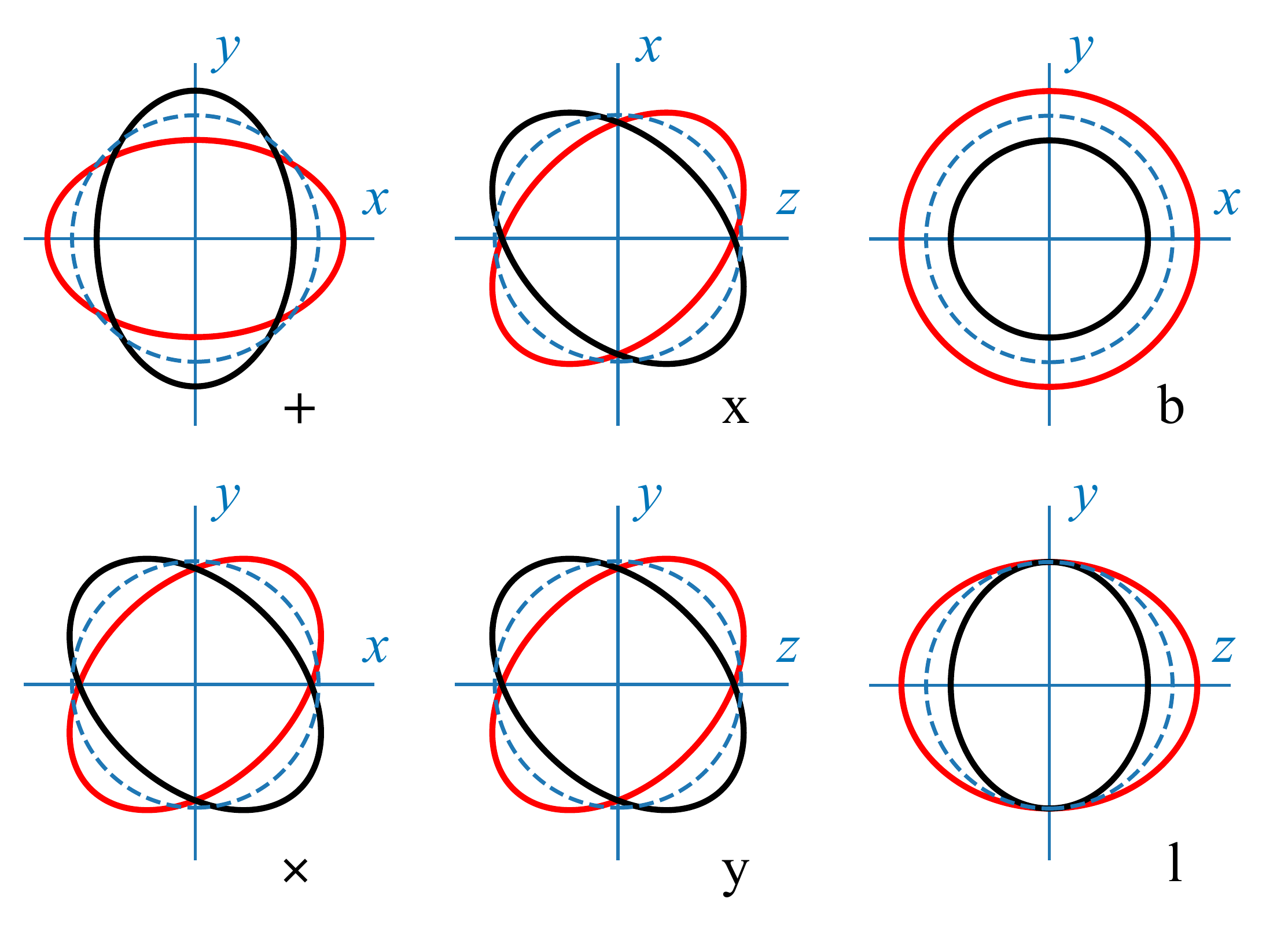}
\caption{The effect of various GW polarizations on a ring of test particles (tensor modes in the left, vector modes in the middle and scalar modes in the right). The wave is always traveling in the $z$ direction. The dashed circles show the original configuration of the test particles before the arrival of the wave and the solid red/black circles and ellipses show the new position of the test particles during the two half cycles of the wave.}
\label{fig:pol_diagram}
\end{figure}

Accurate measurement of the polarizations of GWs provide yet another opportunity to test the predictions of GR. According to GR, GWs have only two independent polarization states --- two transverse quadrupole (or, tensor) modes, while a general metric theory of gravity can admit up to six polarization modes (Fig.~\ref{fig:pol_diagram}). For example, scalar-tensor theories admit two monopole (or, scalar) modes in addition to the tensor modes --- massless scalar-tensor theories admit a transverse scalar (or, breathing) mode, while massive scalar-tensor theories admit both transverse and longitudinal scalar modes~\cite{Will:2014kxa}. More general theories, such as bimetric theories \cite{Schmidt-May:2015vnx}, also admit two dipole (or, vector) modes. 

GW polarizations can be constrained from the observation of long-lived signals from spinning neutron stars~\cite{Isi:2017equ} and stochastic sources~\cite{Callister:2017ocg,Abbott:2018utx,Nishizawa:2009bf} as well as from the observation of transient sources such as compact binary mergers~\cite{Isi:2017fbj,Pang:2020pfz}. While the detection probabilities of spinning neutron stars and stochastic background are uncertain, we are expecting the detection of hundreds to thousands of compact binary signals in the next few years, using ground-based GW detectors such as LIGO~\cite{LIGODet}, Virgo~\cite{VirgoDet}, KAGRA~\cite{KAGRADet} and LIGO-India~\cite{LI-Det}. Note that each of these quadrupole detectors observes only \emph{one} linear combination of these polarizations. The relative strength of each polarization mode in the observed signal in each detector depends on the response of the detector to the specific polarization. It turns out that the detector response to both the scalar modes (breathing and longitudinal modes) are identical, making them completely degenerate~\cite{Chatziioannou:2012rf}. In summary, even in the ideal case, if we want to disentangle all the five non-degenerate polarization modes from the GW data, we need at least five detectors having different orientations. This would be challenging even when upcoming detectors such as KAGRA and LIGO-India join the international GW network, since the two LIGO detectors in Hanford and Livingston are co-aligned with each other and hence measure the \emph{same} linear combination of the polarizations.  

Given the data from a network of GW detectors, we can compare the posterior probabilities of different hypotheses, for example, one hypothesis stating that the polarizations are exactly as predicted by GR, while the alternative hypothesis accommodating the presence of additional modes~\cite{Isi:2017fbj}. Motivated by the limited number of linearly independent detectors to observe the polarization modes, the current probes of the nature of GW polarizations have employed highly simplified hypotheses as alternatives to GR. That is, the alternative hypothesis assumes that the polarizations contain only scalar modes or only vector modes (no tensor modes). The analyses of some of the compact binary merger observations by LIGO and Virgo have concluded that the tensor-only hypothesis is preferred over scalar-only or vector-only hypotheses~\cite{GW170814,GWTC1TestofGR}. 

GWs are gravitationally lensed by intervening matter distributions, such as galaxies and galaxy clusters. Although the gravitational lensing is best understood in GR, alternative theories also predict this effect~(see, e.g.,~\cite{Bekenstein:1993fs}). Recent estimates suggest that a small fraction ($\sim 0.1\%-0.5\%$) of the binary black hole (BBH) signals that we expect to detect using the LIGO-Virgo network will be strongly lensed by intervening galaxies~\cite{Haris:2018vmn, Ng:2017yiu}, producing multiple ``images'' of the signals~\footnote{Galaxy clusters will also cause strong lensing of the GW signals. However, the lensing probability due to clusters is likely to be significantly smaller than that by galaxies~\cite{Smith:2018gle}. However, cluster lensing is expected to produce longer time delays between images, which will further improve the our ability for polarization model selection from lensed GW signals.}. These signals arrive at the detector with relative time delays ranging from several minutes to several weeks~\cite{Haris:2018vmn}. Lensing by galaxies or clusters is very well approximated by geometric optics since the mass scale of the lens is significantly larger than the wavelength of GWs ($GM_\mathrm{lens}/c^2 \gg \lambda_\mathrm{GW}$). Thus, multiple images would correspond to copies of the same signal with a relative magnification and time delay. 

Due to the rotation of the earth, observed signals from multiple images will involve different combinations of the same polarizations. As far as the polarization content is concerned, this is equivalent to observing the same signal with a multiplied number of detectors. For example, if two images of the merger are observed using a three-detector network, this is equivalent to observing the one merger signal with a six-detector network. In this paper, we explore the possibility of constraining the polarization content of GWs using BBH mergers that are strongly lensed by galaxies. We use the Bayesian model selection method proposed by~\cite{Isi:2017fbj} to identify the polarization content of simulated GW signals from binary black holes. We show that strongly lensed GW signals will enable us to constrain the polarization content significantly better than their unlensed counterparts. 

The rest of the paper is organized as follows: Section~\ref{sec:method} summarizes our methodology, providing a brief introduction to the relevant theory, model selection formalism, as well as the details of the numerical simulations. Section~\ref{sec:results} presents the results, while some concluding remarks and future work is discussed in Sec.~\ref{sec:summary}.

\section{Method}
\label{sec:method}

\subsection{GW polarizations}
\label{sec:GWpols}
In the local Lorentz gauge, the spatial components of the metric perturbation $\h_{ij}$ at a given space-time point $\vec x$ can be written in terms of six linearly independent polarization tensors, $\e^A$~\footnote{In this Section, we denote four-vectors by the use of arrows (e.g., $\vec x$) and three-vectors by boldface (e.g., $\bf x$) and tensors by sans serif fonts  (e.g., $\e$). Repeated indices are assumed to be summed over.}
\begin{equation}
\h_{ij}(\vec x) = h_A(\vec x) \, \e^A_{ij}  , \qquad A \in \{+,\times, \mathrm{x,y,b,l}\}
\end{equation}
where, the index $A$ stands for different polarizations: tensor ``plus'' ($+$) and ``cross'' ($\times$) modes, vector ``x'' and ``y'' modes and scalar ``breathing'' (b) and ``longitudinal'' (l) modes; and $h_A$ is the amplitude for polarization $A$. The existence of six independent polarization modes (or, six linearly independent components of the metric perturbation) can be understood in the following way: the full metric perturbation $h_{\mu\nu}$ in four dimensions is symmetric and therefore has ten independent components. However, because of the Lorentz gauge condition, four degrees of freedom are taken away, leaving only six. (GR, in addition, satisfies the transverse-traceless gauge condition which takes away additional 4 degrees of freedom and hence allowing only two tensor polarization modes). 

Further, the polarization tensors can be written in terms of the orthogonal basis vectors $\mathbf{w}_x$, $\mathbf{w}_y$, $\mathbf{w}_z \equiv \mathbf{w}_x \times \mathbf{w}_y$, where $\mathbf{w}_z$ is the GW propagation direction. 
\begin{eqnarray}
\e^{+} &=& \mathbf{w}_{x} \otimes \mathbf{w}_{x}-\mathbf{w}_{y} \otimes \mathbf{w}_{y}  \nonumber \\
\e^{\times} &=& \mathbf{w}_{x} \otimes \mathbf{w}_{y}+\mathbf{w}_{y} \otimes \mathbf{w}_{x} \nonumber \\
\e^{\rm x} &=& \mathbf{w}_{x} \otimes \mathbf{w}_{z}+\mathbf{w}_{z} \otimes \mathbf{w}_{x} \nonumber \\
\e^{\rm y} &=& \mathbf{w}_{y} \otimes \mathbf{w}_{z}+\mathbf{w}_{z} \otimes \mathbf{w}_{y} \nonumber \\
\e^{\rm b} &=& \mathbf{w}_{x} \otimes \mathbf{w}_{x}+\mathbf{w}_{y} \otimes \mathbf{w}_{y} \nonumber \\
\e^{\rm l} &=& \mathbf{w}_{z} \otimes \mathbf{w}_{z}
\label{eq:poltensor}
\end{eqnarray}
A ground-based laser interferometric detector measures a combination of these polarizations by the change in lengths of its perpendicular arms. This response is encoded in the detector tensor $\D$, whose components are given by  
\begin{equation}
\D^{ij} = \frac{1}{2} \, \left( d_x^i d_x^i - d_y^j d_y^j \right)
\label{eq:dettensor}
\end{equation}
where $\d_x$ and $\d_y$ are unit vectors along the detector arms, with a common origin. The strain, $h_I$ measured by detector $I$, is then given as,
\begin{equation}
h_I(t) = \h_{ij}(t,\mathbf{x}_I) \, \D^{ij}_I = h_A(t, \mathbf{x}_I) \, \e^A_{ij} \, \D_I^{ij} = h_A(t,\mathbf{x}_I) \, F^A_I
\label{eq:h_of_t}
 \end{equation}
where, $F^A_I  \equiv \D^{ij}_I \, \e^A_{ij}$, are called the detector antenna pattern functions, which encode the response of the detector $I$ to polarization $A$. Therefore, GW strain measured at the detector $I$ can be written as the linear combination of polarization amplitudes multiplied with the corresponding antenna pattern functions. Expanding, $F_A$ using Eqs.~\eqref{eq:poltensor} and \eqref{eq:dettensor}: 
\begin{eqnarray}
F_{+} & = & \frac{1}{2}\left[\left(\mathbf{w}_{x} \cdot \mathbf{d}_{x}\right)^{2}-\left(\mathbf{w}_{x} \cdot \mathbf{d}_{y}\right)^{2}-\left(\mathbf{w}_{y} \cdot \mathbf{d}_{x}\right)^{2}+\left(\mathbf{w}_{y} \cdot \mathbf{d}_{y}\right)^{2}\right] \nonumber \\
F_{\times} &= & \left(\mathbf{w}_{x} \cdot \mathbf{d}_{x}\right)\left(\mathbf{w}_{y} \cdot \mathbf{d}_{x}\right)-\left(\mathbf{w}_{x} \cdot \mathbf{d}_{y}\right)\left(\mathbf{w}_{y} \cdot \mathbf{d}_{y}\right)   \nonumber \\
F_{\mathrm{x}} &= &\left(\mathbf{w}_{x} \cdot \mathbf{d}_{x}\right)\left(\mathbf{w}_{z} \cdot \mathbf{d}_{x}\right)-\left(\mathbf{w}_{x} \cdot \mathbf{d}_{y}\right)\left(\mathbf{w}_{z} \cdot \mathbf{d}_{y}\right)  \nonumber \\
F_{\mathrm{y}} & = & \left(\mathbf{w}_{y} \cdot \mathbf{d}_{x}\right)\left(\mathbf{w}_{z} \cdot \mathbf{d}_{x}\right)-\left(\mathbf{w}_{y} \cdot \mathbf{d}_{y}\right)\left(\mathbf{w}_{z} \cdot \mathbf{d}_{y}\right)  \nonumber \\
F_{\mathrm{b}} & = &\frac{1}{2}\left[\left(\mathbf{w}_{x} \cdot \mathbf{d}_{x}\right)^{2}-\left(\mathbf{w}_{x} \cdot \mathbf{d}_{y}\right)^{2}+\left(\mathbf{w}_{y} \cdot \mathbf{d}_{x}\right)^{2}-\left(\mathbf{w}_{y} \cdot \mathbf{d}_{y}\right)^{2}\right]  \nonumber \\
F_{\mathrm{l}} & = & \frac{1}{2}\left[\left(\mathbf{w}_{z} \cdot \mathbf{d}_{x}\right)^{2}-\left(\mathbf{w}_{z} \cdot \mathbf{d}_{y}\right)^{2}\right]
\label{eq:antenna_patt}
\end{eqnarray}
Evaluating these antenna pattern functions at a particular detector involves projecting the polarization tensors into the detector frame. This projection depends on the direction from which GW arrives for a particular detector and hence the sky-location of the GW source. We choose to describe the source using the equatorial coordinate system, in terms of right ascension $\alpha$ and declination $\delta$. Additionally, antenna pattern functions also depend on polarization angle, $\psi$, which is due to the rotational freedom of the orthonormal vectors ($\mathbf{w}_x , \mathbf{w}_y$) about the propagation direction $\mathbf{w}_z$. Further, the detector response and hence the antenna pattern functions also depend on time due to the rotation of the earth. Thus, the antenna pattern functions of the detector $I$ for the polarization mode $A$ can, in general, be written as $F_I^A(\alpha, \delta, \psi, t)$. The antenna pattern functions for tensor polarizations are computed using the \textsc{PyCBC} software package~\cite{pycbc}.  We added in it the antenna response to the scalar and vector polarizations using Eq.\eqref{eq:antenna_patt}. 

For three detector network, we have three strain measurements giving three different linear combinations of polarizations, as given by Eq.\eqref{eq:h_of_t}. However, due to strong lensing, multiple copies of the same GW signal would arrive at each detector with time delay $\Delta t$ of minutes to weeks. Due to the rotation of the earth, the antenna patterns during the arrival of, say two images, $F_I^A(\alpha, \delta, \psi, t)$ and $F_I^A(\alpha, \delta, \psi, t+\Delta t)$ can be considerably different from each other. This is equivalent to observing one signal with a six-detector network. 

According to GR, in the geometrical optics limit, polarization tensors are parallelly propagated along the null geodesics, implying that lensing does not change the polarization content of a GW. We assume this to be true in alternative theories also. As long as the metric perturbation follow a source-free wave equation, the polarization tensor should be conserved along the GW propagation.

\subsection{Model selection of polarizations}

Bayesian model selection allows us to assign posterior probabilities for various hypotheses pertaining to the observed data. We formulate the polarization content of GWs as different Bayesian hypotheses. For e.g., GR is denoted as $\Ht$, as the theory only predicts tensor modes. The hypothesis that GWs contain only scalar (vector) modes is denoted as $\Hs$ ($\Hv$). Following~\cite{Isi:2017equ}, we assume that the waveforms in $\Hs$ and $\Hv$ are the same as in $\Ht$; the only change is in the antenna pattern functions. We do this as the BBH waveforms for alternative theories are presently not known. If available in the future they can be included in the same formalism.

Given the set of data $\{d\}$ from a network of detectors, the marginalized likelihood (or, Bayesian evidence) of the hypothesis $\H_p$ can be computed by 
\begin{equation}
P(\{d\}|\H_p) = \int d\btheta P(\btheta) \, P(\{d\}|\btheta, \H_p),
\label{eq:marg_likelihood}
\end{equation}
where $\btheta$ is a set of parameters that describe the signal under hypothesis $\H_p$ (including the masses and spins of the compact objects in the binary, location and orientation of the binary and the arrival time and phase of the signal), $P(\btheta)$ is the prior distribution of $\btheta$ (which we take to be independent of $\H_p$), and $P(\{d\}|\btheta, \H_p)$ is the likelihood of the data $\{d\}$, given the parameter vector $\btheta$ and hypothesis $\H_p$. Given the hypothesis $\H_p$ and data $\{d\}$, we can sample and marginalize the likelihood over the parameter space using an appropriate stochastic sampling technique such as Nested Sampling~\cite{skilling2006}. 

Bayesian model selection allows us to compare multiple hypotheses. For e.g., the odds ratio $\Ots$ is the ratio of the posterior probabilities of the two hypotheses $\Ht$ and $\Hs$. When $\Ots$ is greater than one then hypothesis $\Ht$ is preferred over $\Hs$ and vice versa. Using Bayes theorem, the odds ratio can also be written as the product of the ratio of the prior odds $\Pts$ of the hypotheses and the likelihood ratio, or Bayes factor $\Bts$: 
 \begin{equation}
\Ots := \frac{P(\Ht|\{d\})}{P(\Hs|\{d\})} = \frac{P(\Ht)}{P(\Hs)} \times \frac{P(\{d\}|\Ht)}{P(\{d\}|\Hs)} = \Pts \times \Bts 
\end{equation}
Since GR has been tested well in a variety of settings, our prior odds are going to be highly biased towards tensor-only modes, i.e., $\Pts \gg 1$. Hence, in order to claim evidence of non-tensor modes the corresponding Bayes factor supporting the alternative hypothesis has to be very large. Since the Bayes factor is the only quantity that is derived from data, for the rest of the paper, we focus on the Bayes factor. The Bayes factor from multiple, uncorrelated events $d^{(i)}$ can be combined as 
\begin{equation}
\Bts =  \prod_i \Bts^{(i)},
\label{eq:combined_BF_unlensed}
\end{equation}
where
\begin{equation}
\Bts^{(i)} = \frac{P(\{d^{(i)}\}|\Ht)}{P(\{d^{(i)}\}|\Hs)}
\label{eq:ts_bayes_factor_event_i}
\end{equation}
is the Bayes factor obtained from the $i$th event. 

\subsection{Model selection of polarizations using lensed GW events}
\label{sec:model_sel_lens}

When multiple GW events are produced by the strong lensing of a signal BBH merger, these events cannot be treated as uncorrelated events. Here we derive the Bayes factor between different polarization hypotheses $\H_p$ using multiple lensed images of the same merger. For simplicity, we will consider only two lensed images. However, the same formalism can be extended to more than two images also. Strong lensing of GWs from BBHs is expected to be dominated by galaxy lenses. Lensing by galaxies and galaxy clusters can be treated in geometric optics regime (wavelength of GWs significantly smaller than the mass scale of the lens). In this regime, lensing does not change the frequency profile of the waveform, and hence multiple images, arriving at the detector at different times, differ from each other only by a relative magnification and a constant phase shift\footnote{{Note that this is valid only for the quadrupole mode GW signals from non-precessing binaries, which is the case we are considering in this paper.}}~\cite{Dai:2017huk,Blandford:1986zz}. Hence the parameters describing the waveform, except for the luminosity distance (which is degenerate with the lensing magnification) and the time and phase at coalescence will be common between the two images. 

Now consider the GW signals $d^{(1)}$ and $d^{(2)}$ produced by the strong lensing of a BBH merger. The Bayes factor between two polarization hypotheses $\Ht$ and $\Hs$ can be written as 
 \begin{eqnarray}
\Bts &:=& \frac{P(\{d^{(1)}, d^{(2)}\}|\Ht)}{P(\{d^{(1)}, d^{(2)}\}|\Hs)} \\
     &=& \frac{\int d\btheta_c \, P(\btheta_c) P(\{d^{(1)}\}|\btheta_c, \Ht) \, P(\{d^{(2)}\}|\btheta_c, \Ht)} {\int d\btheta_c \, P(\btheta_c) P(\{d^{(1)}\}|\btheta_c, \Hs) \, P(\{d^{(2)}\}|\btheta_c, \Hs)},
\label{eq:HtHs_bf_lens}
\end{eqnarray}
where $\btheta_c$ is the vector of common parameters as the signals come from the same merger. Note that the probability distributions are marginalized over all the parameters except $\btheta_c$. Using the Bayes theorem, the likelihoods $P(\{d^{(i)}\}|\btheta_c, \H_p)$ can be written in terms of the posteriors $P(\btheta_c | \{d^{(i)}\}, \H_p)$ as 
\begin{equation}
P(\{d^{(i)}\}|\btheta_c, \H_p) = \frac{P(\btheta_c | \{d^{(i)} \}, \H_p) \, P(\{d^{(i)}\}, \H_p)}{P(\btheta_c)},
\end{equation}
where $P(\{d^{(i)}\}| \H_p)$ is the marginal likelihood of the hypothesis $\H_p$ defined in Eq.\eqref{eq:marg_likelihood}. Using this, Eq.\eqref{eq:HtHs_bf_lens} can be rewritten as 
\begin{eqnarray}
\Bts = \Bts^{(1)} \, \Bts^{(2)} \, \frac{\Blu|\Ht}{\Blu|\Hs},
\label{eq:HtHs_bf_lens2}
\end{eqnarray}
where $\Bts^{(1)}$ and $\Bts^{(2)}$ are the Bayes factors of the polarization hypotheses obtained from event 1 and 2, respectively [see Eq.\eqref{eq:ts_bayes_factor_event_i}], while $\Blu$ is the lensing Bayes factor defined in~\cite{Haris:2018vmn}. That is, 
\begin{eqnarray}
\Blu|\H_p = \int \frac{d\btheta_c \, P(\btheta_c|\{d^{(1)}\}, \H_p)  \, P(\btheta_c|\{d^{(2)}\}, \H_p)}{P(\btheta_c)}.
\label{eq:blu}
\end{eqnarray}
This is the Bayes factor between a different set of two hypotheses $\Hl$ and $\Hu$, which are different from the hypotheses on polarization content. The lensing hypothesis $\Hl$ states that this pair of events are lensed images of the same merger, while the unlensed hypothesis $\Hu$ states that these are unrelated events. It can be seen from Eq.\eqref{eq:blu} that $\Blu|\H_p$ is the prior-weighted inner product of the posteriors of the common parameters $\btheta_c$ obtained from the two images. These posteriors are computed assuming the polarization hypothesis $\H_p$. Note that these posteriors, and hence the lensing Bayes factor can be computed assuming different hypotheses for the polarization content $\H_p = \{\Ht, \Hs, \Hv\}$.

From Eq.\eqref{eq:HtHs_bf_lens2} it is evident that, given a pair of lensed events, the combined Bayes factor between the two polarization hypotheses is the product of the Bayes factors computed from the individual events multiplied by an extra term $\frac{\Blu|\Ht}{\Blu|\Hs}$, which we call the {\it posterior overlap ratio}. We can do parameter estimation for the individual events assuming different polarization hypotheses $\H_p$ to get the posteriors and marginal likelihoods (evidence) from each event. Later, from these posteriors, we compute the overlap factors $\Blu$ of the two events. If the posterior overlaps using the correct polarization hypothesis, say $\Ht$, is larger than the same using the wrong hypothesis, say $\Hs$, (that is, if ${\Blu|\Ht} > {\Blu|\Hs}$), then the combined Bayes factor $\Bts$ of the two lensed events will be larger than the same computed from two unlensed events with the same individual Bayes factors $\Bts^{(1)}$ and $\Bts^{(2)}$. This suggests that lensed events can improve our ability to identify the right polarization hypothesis, with an improvement factor given by the posterior overlap ratio $\frac{\Blu|\Ht}{\Blu|\Hs}$. 

\subsection{Simulations}

\begin{figure}[tbh]
\includegraphics[width=0.95\columnwidth]{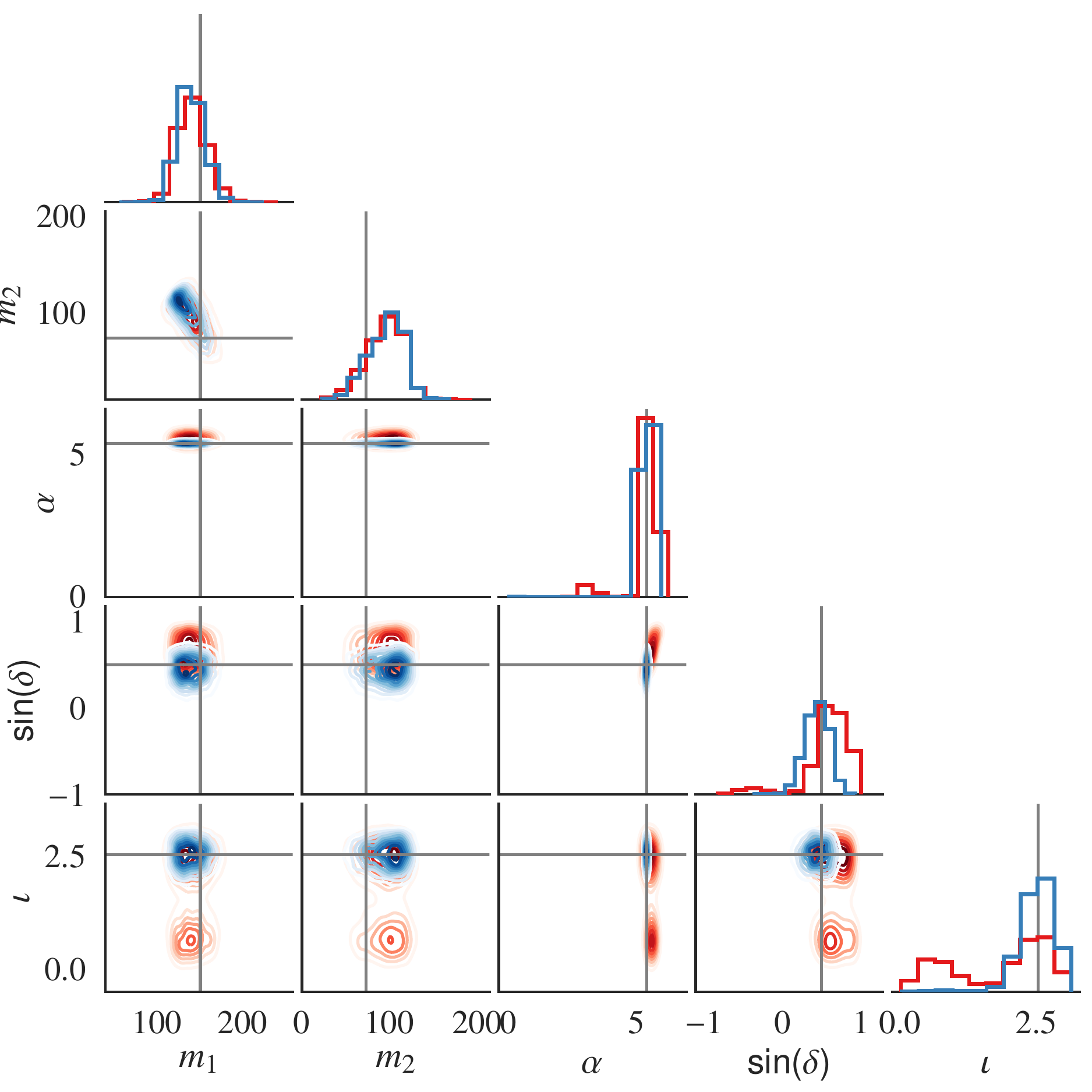}
\caption{Marginalized posteriors of the parameters $m_1, m_2, \alpha, \sin \delta, \iota$ estimated from a lensed pair of tensor injections with tensor recovery (i.e., $\Ht^{[I]}-\Ht^{[R]}$ case). Gray lines show the injected values. Note that the posteriors estimated from the two images are overlapping and are consistent with the injected values.}
\label{fig:T-T}
\end{figure}

\begin{figure}[tbh]
\includegraphics[width=0.95\columnwidth]{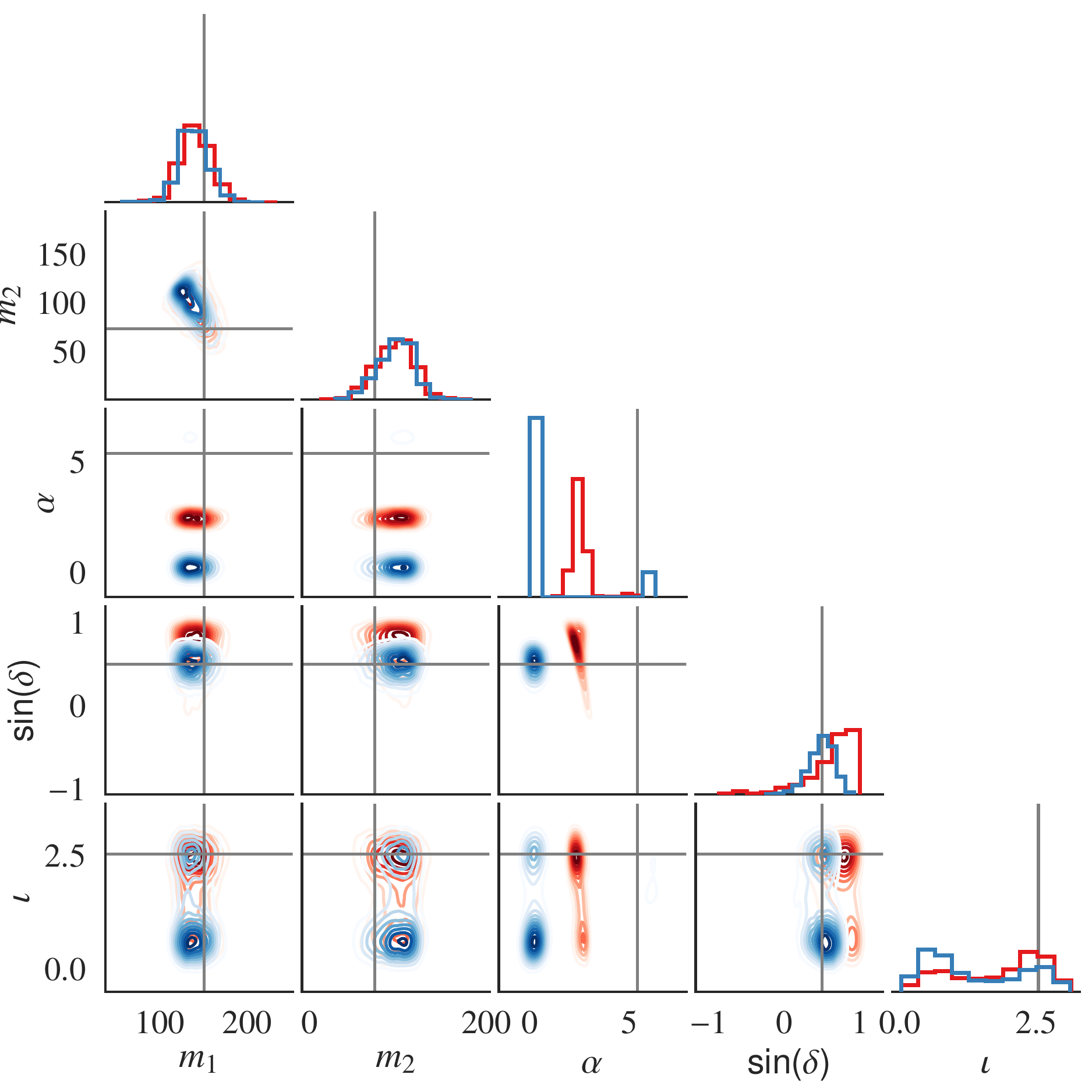}
\caption{Same as Fig.~\ref{fig:T-T}, except that that the injection is performed using the tensor polarization model while parameter estimation is performed using the vector model (i.e., $\Ht^{[I]}-\Hv^{[R]}$ case). Note that the posteriors of the extrinsic parameters $\alpha, \sin \delta, \iota$, estimated from the two images, are not always overlapping and are not always consistent with the injected values.}

\label{fig:T-V}
\end{figure}	
\begin{figure}[tbh]
\includegraphics[width=0.95\columnwidth]{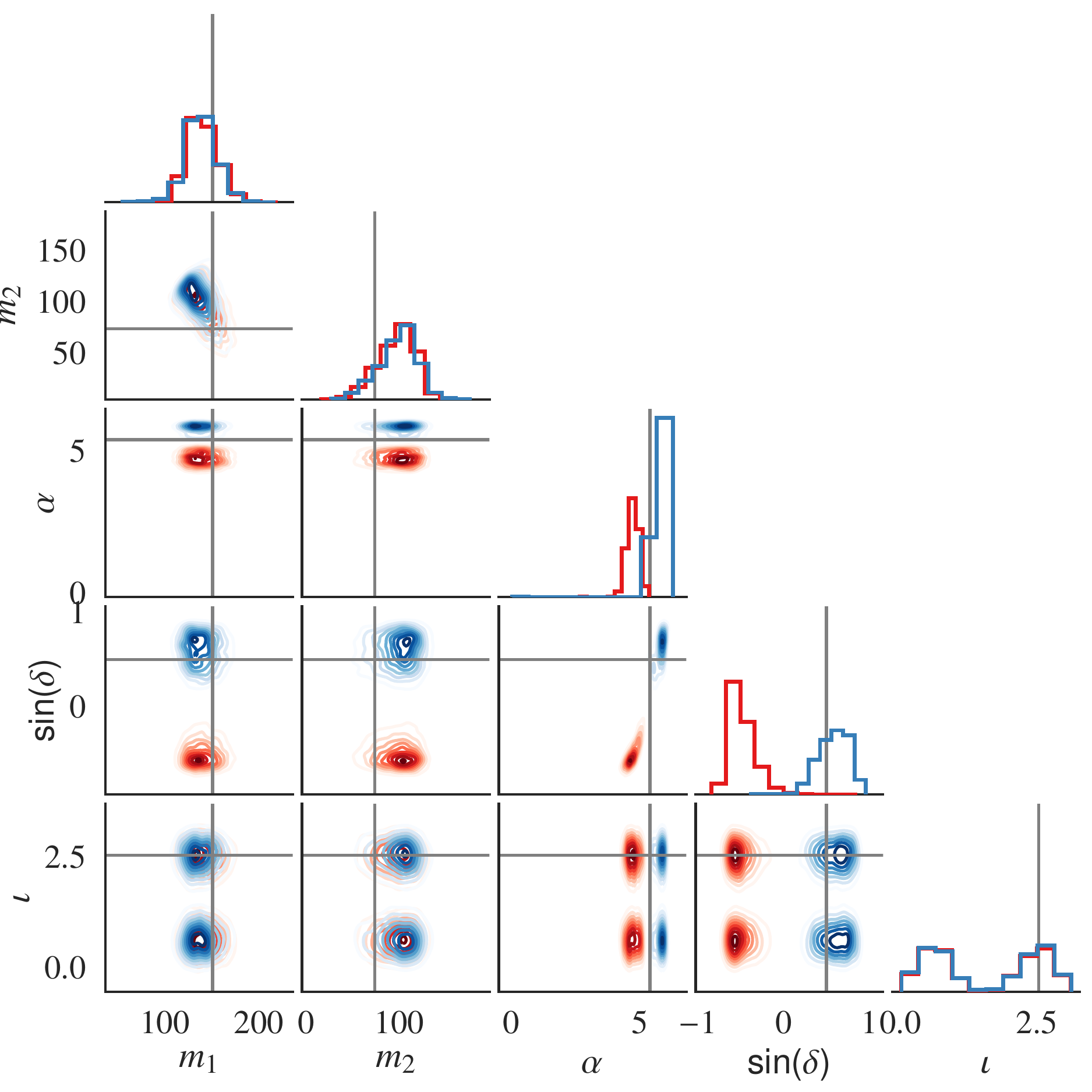}
\caption{Same as Fig.~\ref{fig:T-V}, except that the injection is performed using the tensor polarization model while parameter estimation is performed using the scalar model (i.e., $\Ht^{[I]}-\Hs^{[R]}$ case). Note that the posteriors of the extrinsic parameters $\alpha, \sin \delta, \iota$, estimated from the two images, are not always overlapping and are not always consistent with the injected values.}
\label{fig:T-S}
\end{figure}	

Since there are no strong lensed events detected till now by the current detectors~\cite{Hannuksela:2019kle,Singer:2019vjs}~\footnote{We note that, during the late states of the preparation of this paper, \cite{Dai:2020tpj} has identified an unusual lensing candidate in the data of the second observing run of LIGO and Virgo.}, we use a simulated catalog of lensed BBH events presented in~\cite{Haris:2018vmn} to study the efficacy of polarization recovery. Simulations in~\cite{Haris:2018vmn} generated the observed parameters for different pairs of lensed BBH merger signals, assuming a well-motivated distribution of lens and source properties~\cite{Haris:2018vmn}. In the geometric optics regime, for a particular event, intrinsic parameters like the black holes' redshifted masses ($m_1$ and $m_2$) and spins as well as the extrinsic parameters like the sky-location ($\alpha$ and $\delta$) and orientation ($\iota$ and $\psi$) of the binary remain the same for the multiple images. Similarly, the coalescence (orbital) phase $\phi_0$ estimated from the two images should be consistent, apart from a possible constant shift of $n\, \pi/4$, where $n$ is an integer~\cite{Dai:2017huk}~\footnote{Since we assume that the hypothesis that the given pair of events are lensed copies of the same merger has been established using prior analysis, we would know the value of this constant phase shift between the two signals.}. However, the strain amplitude, magnified due to strong lensing and is degenerate with the luminosity distance ($d_L$) is different. Apart from this, the observed time of coalescence ($t_0$) of the two signals is different due to the time delay between the arrival of the two signals (ranging from several minutes to several weeks). Therefore, except $d_L$  and $t_0$, all the other parameters can contribute to our common parameters vector $\btheta_c$. We limit ourselves to non-spinning binaries and perform Bayesian parameter estimation of the following parameters: $m_1, m_2, \alpha, \delta, d_L, t_0, \phi_0, \iota, \psi$. However, in order to compute the posterior overlap ratio, we consider only the following parameters: {$m_1, m_2, \alpha, \delta, \iota$. That is, the posteriors are marginalized over all other parameters. This is based on our empirical observation that this choice of parameters typically provide the largest values of the posterior overlap ratios.} 

From the simulated parameters of the lensed events, we generate GW signals at different detectors for each polarization hypothesis --  tensor $\Ht$, vector $\Hv$, and scalar $\Hs$, using the corresponding antenna pattern functions (see Sec.~\ref{sec:GWpols}). The model signals are generated using the antenna patterns of the corresponding polarizations, but always assuming that the time evolution of the simulated waveform always follow the GR waveform. That is, $h_\mathrm{s}(t) = h_\mathrm{x}(t)  = h_+(t) $ and $h_\mathrm{l}(t)  = h_\mathrm{y}(t) = h_\times(t)$.

For these simulations, we consider a three detector network consisting of two US-based Advanced LIGO detectors located in Hanford, WA and Livingston, LA and the Advanced Virgo detector located in Pisa, Italy. The LIGO detectors were assumed to have their design sensitivity with the power spectral density (PSD) given in~\cite{aLIGOSensitivity} while the Virgo detector was assumed to have the PSD given in~\cite{AvirgoBaseline}. GW signals were simulated using the \textsc{IMRPhenomPv2} waveform approximant~\cite{Hannam:2013oca,Husa_2016IMRPhenomD,Khan_2016IMRPhenomD} coded in the \textsc{LALSimulation} module of the \textsc{LALSuite} software package~\cite{lalsuite}. {We select $\sim 100-150$ injections crossing a threshold of 8 for the network SNR.} Once we have the injections, we use the Dynamic Nested Sampling~\cite{Dynesty:2020} implementation (\textsc{Dynesty}) in \textsc{PyCBCInference} package \cite{Biwer:2018osg,pycbc} to compute the posteriors of the binary parameters and the evidences of each polarization hypothesis $\H_p$. We have three simulated (injection) models $\Ht^{[I]}, \Hs^{[I]}, \Hv^{[I]}$  and three recovery models $\Ht^{[R]}, \Hs^{[R]}, \Hv^{[R]}$, allowing us to analyze the nine combinations of injection sets:  $\Ht^{[I]}-\Ht^{[R]}$, $\Ht^{[I]}-\Hs^{[R]}$, $\Ht^{[I]}-\Hv^{[R]}$, $\Hs^{[I]}-\Ht^{[R]}$, $\Hs^{[I]}-\Hs^{[R]}$, $\Hs^{[I]}-\Hv^{[R]}$, $\Hv^{[I]}-\Ht^{[R]}$, $\Hv^{[I]}-\Hs^{[R]}$ and $\Hv^{[I]}-\Hv^{[R]}$. 

We use the standard Gaussian likelihood model for estimating the posteriors of the parameters under different polarization hypotheses~(see, e.g.,~\cite{Veitch:2014wba}). For simplicity, no noise is added to the simulated signals. Further, we considered only non-spinning binaries. Thus, the likelihood is computed over the following parameters $m_1, m_2, \alpha, \delta, d_L, \iota, \psi, \phi_0, t_0$. {We use uniform priors in component masses of the binary {($m_1, m_2 \in [3,500 ] M_\odot$)}, isotropic sky location (uniform in $\alpha, \sin \delta$) and orientation (uniform in $\cos \iota, \phi_0$), uniform in polarization angle $\psi$, and a volumetric prior $\propto d_L^2$ on luminosity distance.} Finally, the posteriors are marginalized over all the parameters except the ones that we consider for calculating the posterior overlaps, i.e., $m_1, m_2, \alpha, \delta, \iota$. 

As one would anticipate, the true (injected) parameters are recovered when the injection and recovery model are the same. As an example, Fig.~\ref{fig:T-T} shows the estimated posterior distributions when the injections and recovery are performed using the same tensor hypothesis (i.e., the $\Ht^{[I]}-\Ht^{[R]}$ combination). In contrast, when parameter estimation on the tensor injection is performed using vector and scalar hypotheses, the intrinsic parameters (primary and secondary masses) are still recovered well, whereas extrinsic parameters such as the sky location and orientation are not recovered well (see Figs.~\ref{fig:T-V} and \ref{fig:T-S}). This is due to the fact that the recovery of the extrinsic parameters heavily depends	 on the antenna pattern functions, which are different for the injection and recovery models. 

Further, note that sky location posteriors ($\alpha$, $\sin \delta$) of the lensed pairs overlap well with the tensor model (Fig. \ref{fig:T-T}) and not so well with vector and scalar models  (Figs.~\ref{fig:T-V} and \ref{fig:T-S} show the $\Ht^{[I]}-\Hv^{[R]}$ and $\Ht^{[I]}-\Hs^{[R]}$ combinations, respectively). As a result, in general, we would expect that the lensing Bayes factor $\Blu$ will be larger for the tensor model. This is quantified in the next section.  
\section{Results}
\label{sec:results}

\begin{figure*}[tbh]
\includegraphics[width=0.685\columnwidth]{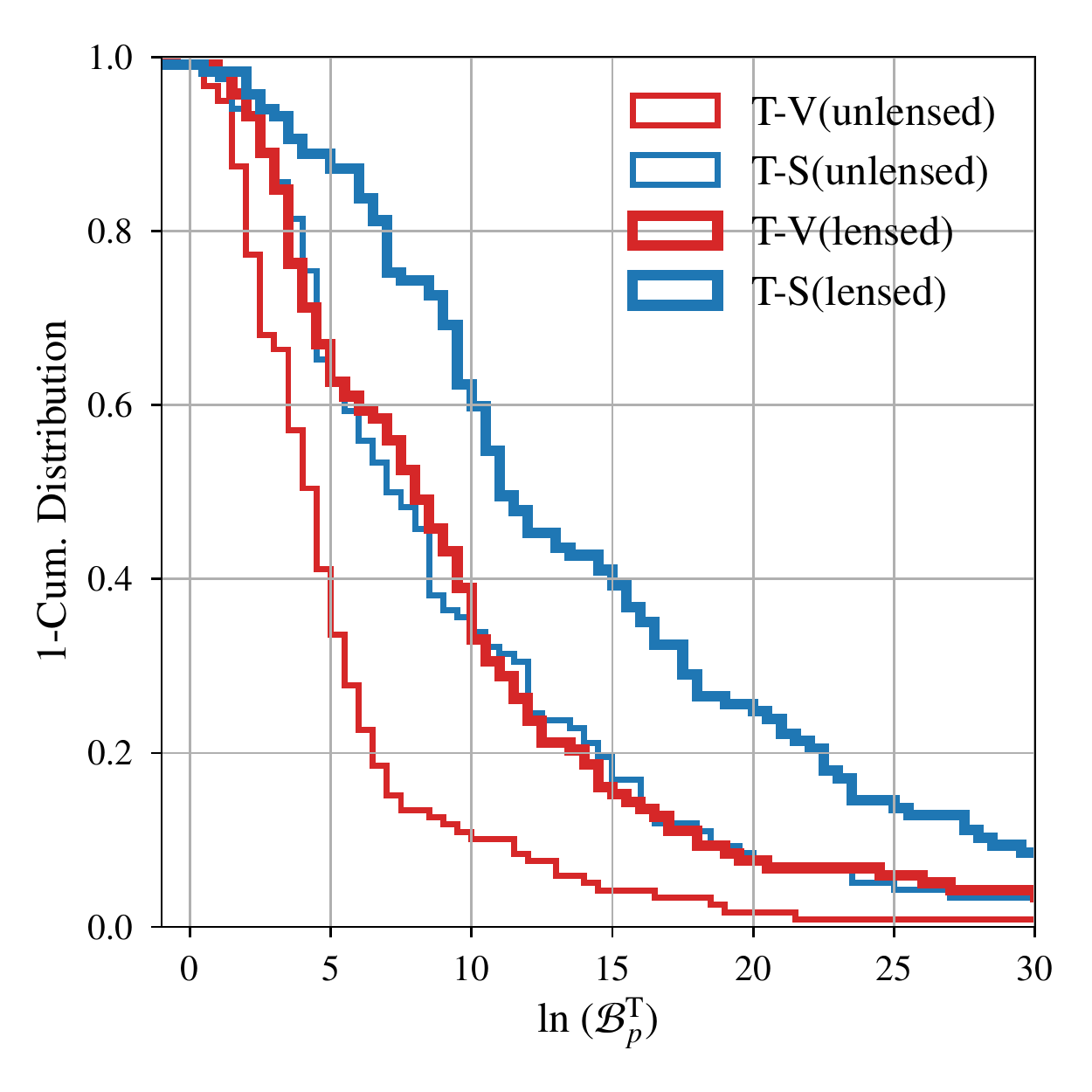} 
\includegraphics[width=0.685\columnwidth]{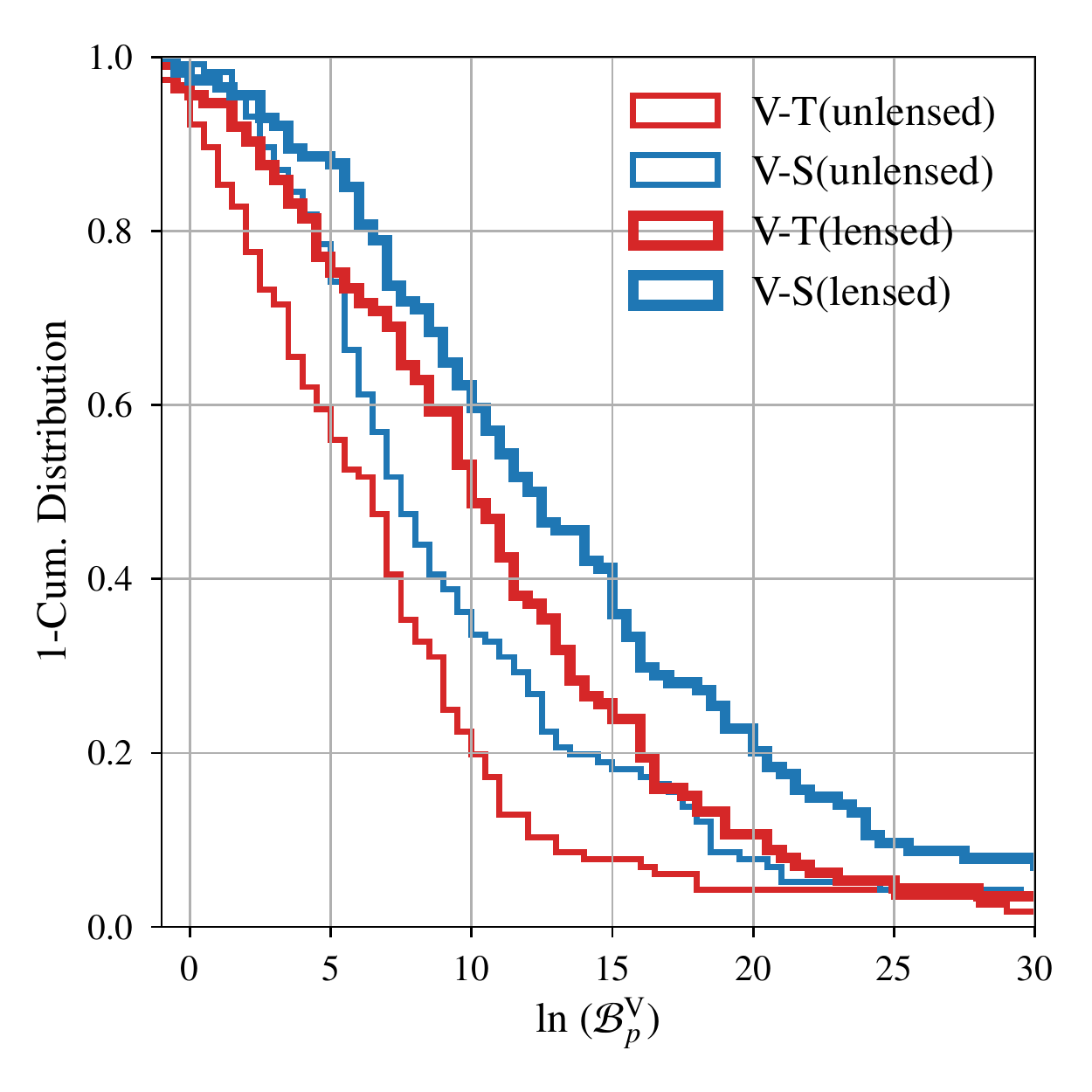} 
\includegraphics[width=0.685\columnwidth]{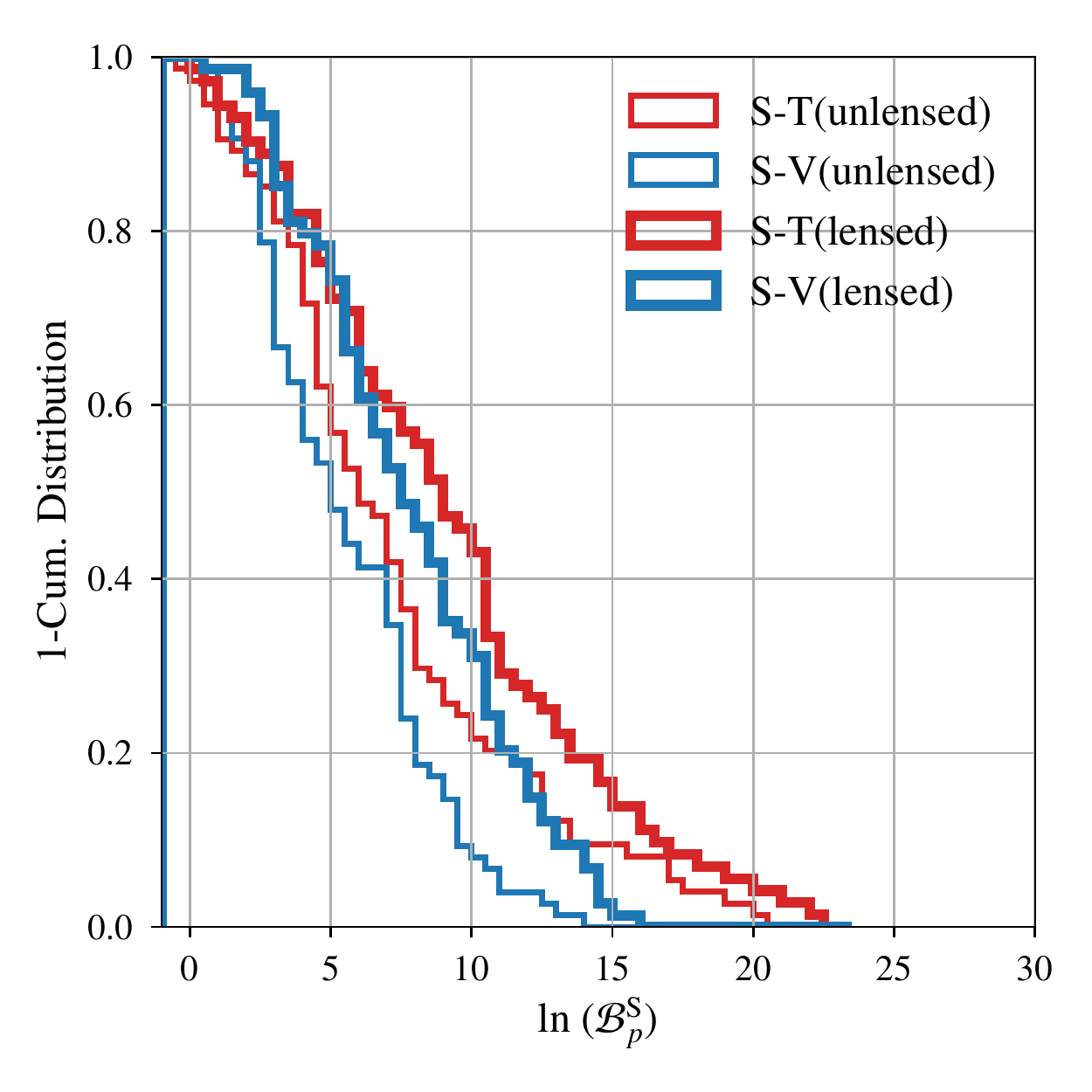} 
\includegraphics[width=0.685\columnwidth]{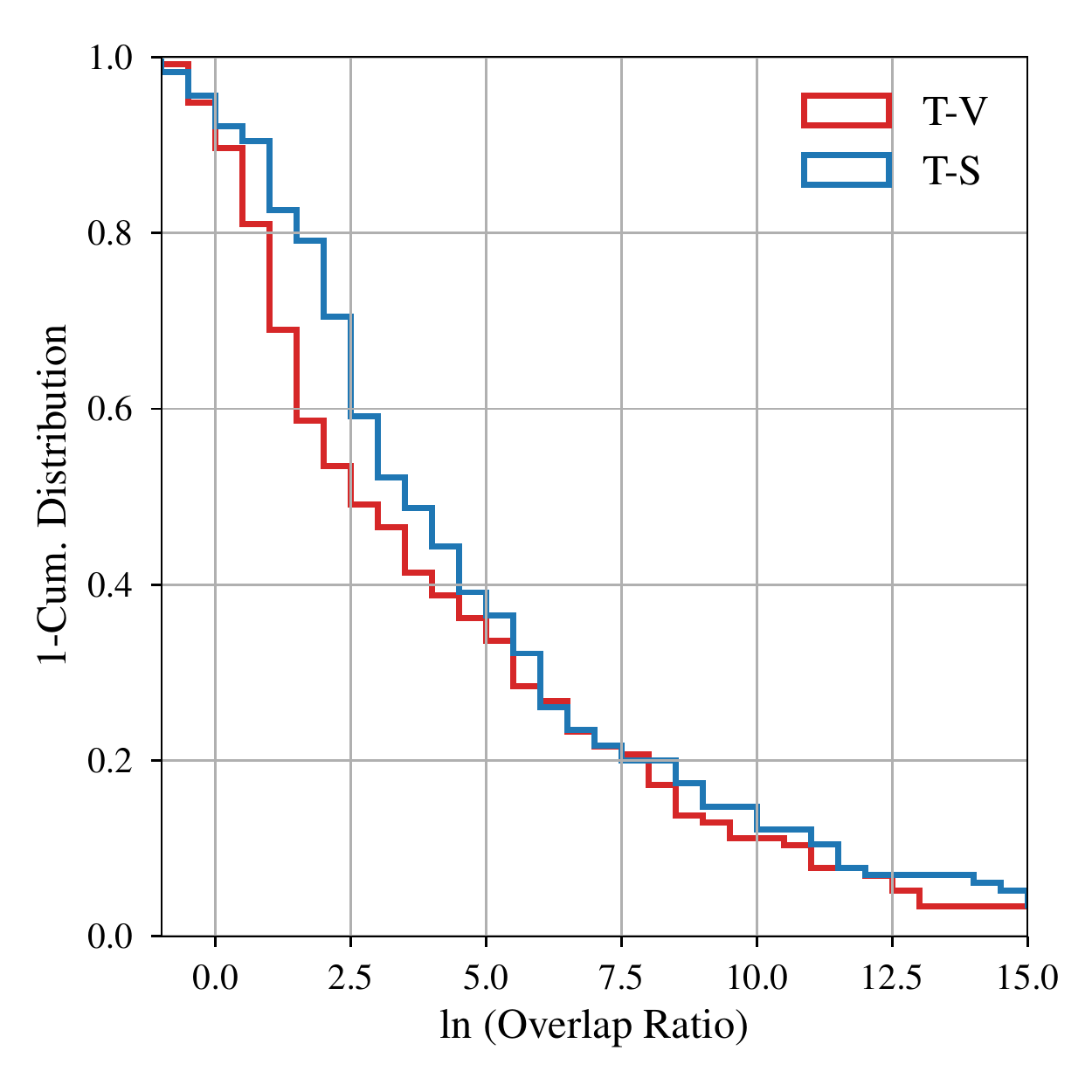}
\includegraphics[width=0.685\columnwidth]{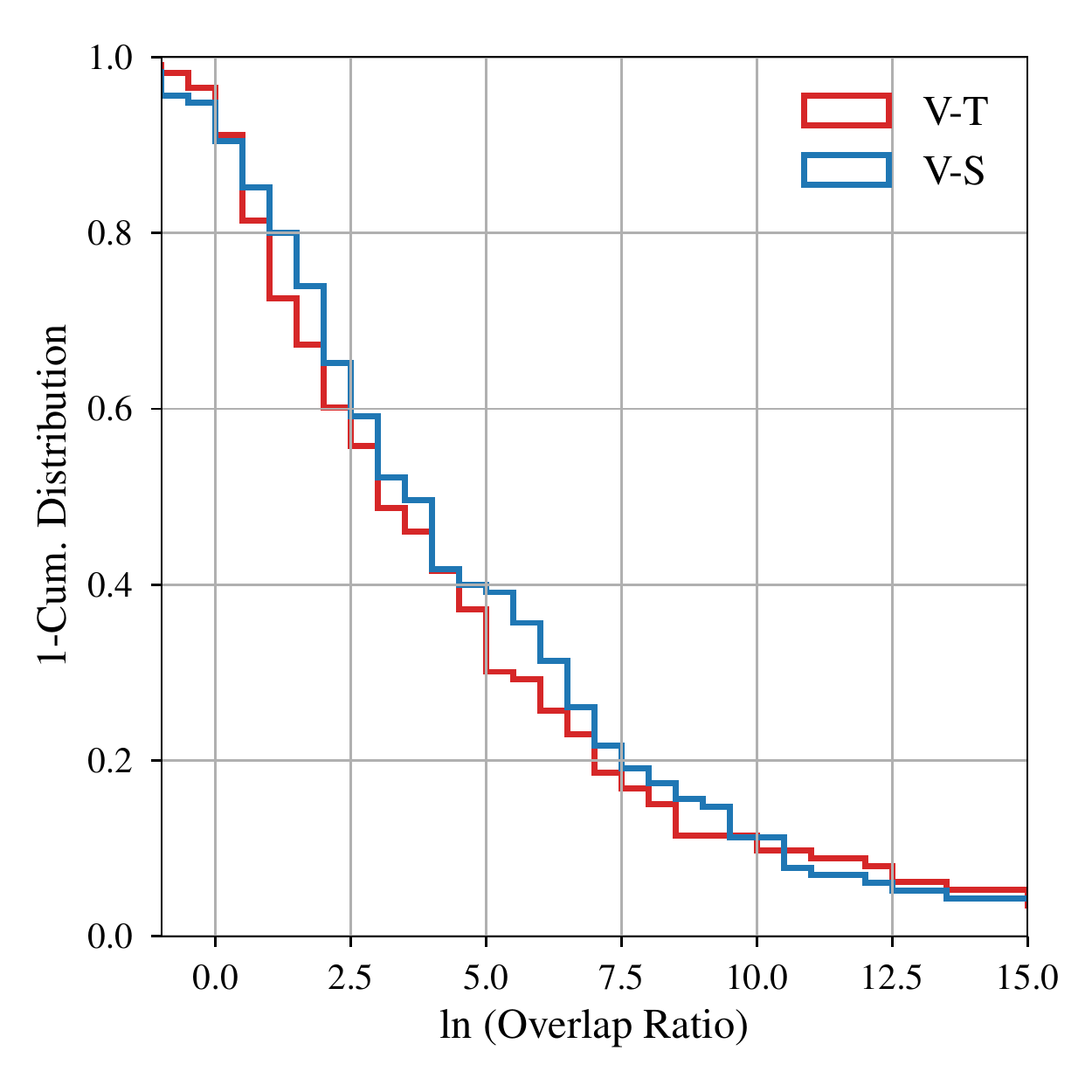}
\includegraphics[width=0.685\columnwidth]{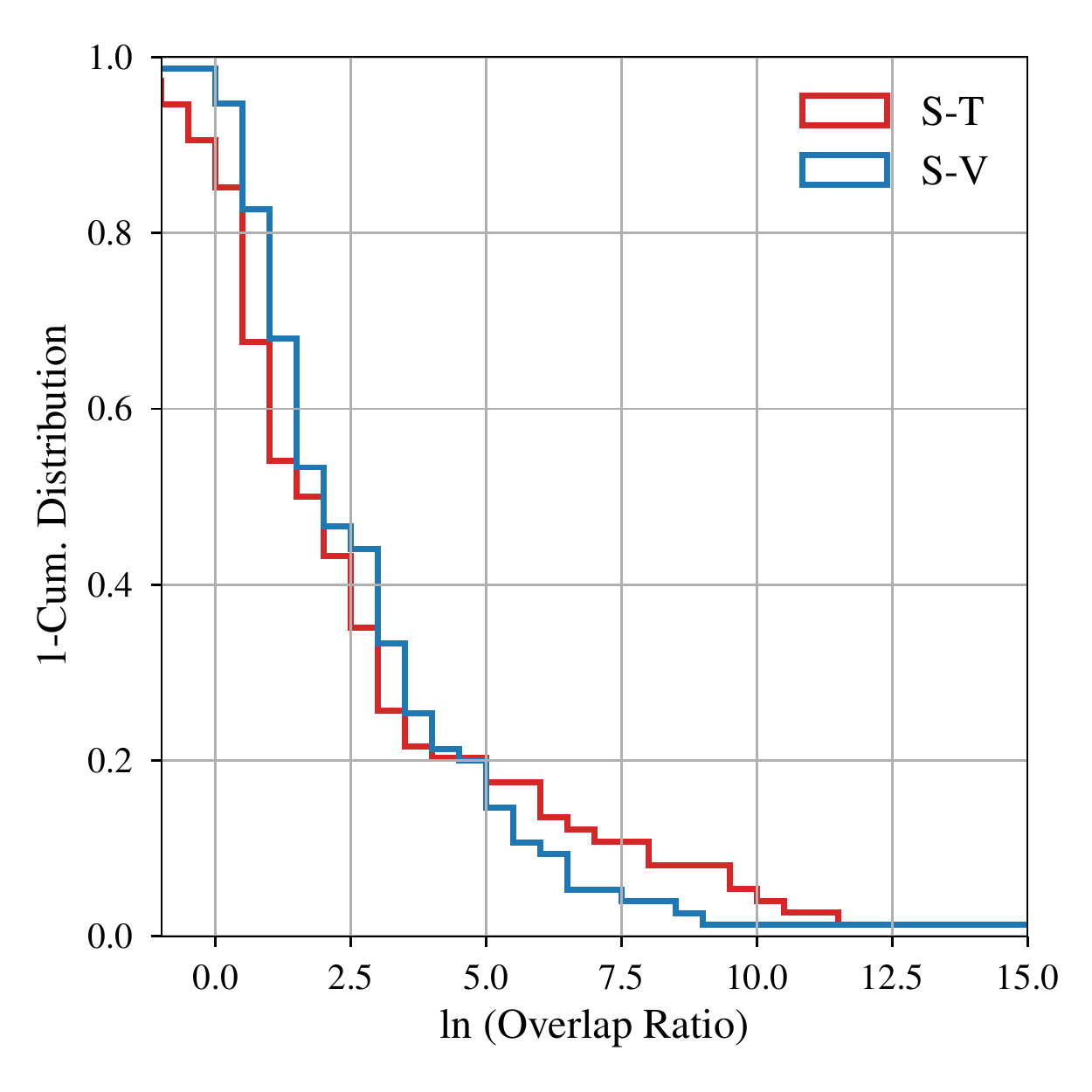}
\caption{\emph{Top panels:} Distribution of the Bayes factor between the ``right'' and ``wrong'' polarization hypotheses estimated from pairs of events (lensed or unlensed). Events are simulated assuming the tensor polarization hypothesis $\Ht$ (left panel), vector polarization hypothesis $\Hv$ (middle panel) as well as scalar polarization hypothesis $\Hs$ (right panel). Each plot shows the distribution of the Bayes factors between the right and wrong polarization hypotheses (for e.g., T-V in the legends denote the Bayes factor $\Btv$). Note that the Bayes factors for the lensed pairs are almost always greater than the same computed from unlensed events with the same parameters. 
\emph{Bottom panels:} Corresponding distribution of the ratios of the overlap factors $\Blu$ assuming the ``right'' and ``wrong'' polarization hypotheses. Note that the overlap ratio is greater than 1 for $85-95\%$ of events, suggesting that lensing improves the Bayes factors of the right hypothesis.}
\label{fig:overlap_ratio_dist}
\end{figure*}

Our aim is to quantify how well pairs of GW signals produced by the strong lensing of BBH mergers improve our ability to distinguish between polarization models, as compared to pairs of unrelated signals with similar strengths. If the two signals are unrelated (i.e., unlensed events) then the combined Bayes factor will just be the product of individual Bayes factors [Eq.\eqref{eq:combined_BF_unlensed}]. On the other hand, if the two events are lensed images of the same merger, then the combined Bayes factor is the product of the individual Bayes factors and additional factor, $\frac{\Blu|\H_\mathrm{right}}{\Blu|\H_\mathrm{wrong}}$, which we call the posterior overlap ratio [Eq.\eqref{eq:HtHs_bf_lens2}]. If the posterior overlaps ratio is greater than one then for the correct polarization hypothesis, this would show that lensing improves our ability to identify the right polarization hypothesis. 

Figure~\ref{fig:overlap_ratio_dist} (top panels) shows the distribution of the polarization Bayes factors from the simulated GW signals. The fact that Bayes factors are almost always greater than 1 (log Bayes factors $>$ 0) suggests that the right polarization hypothesis is almost always preferred. Note that, overall, the lensed Bayes factors are greater than unlensed ones, showing that the strong lensing improves the polarization models selection. This is also evident from the distribution of the posterior overlap ratios (bottom panel of Fig.~\ref{fig:overlap_ratio_dist}). Note that  the overlap ratios are greater than 1 (log overlap ratio $>$ 0) for {$\sim 85-95 \%$} of the simulated events.  The median value of the overlap ratio is {$\sim 2-3$}, which means that for more than 50\% of the events lensing improves the polarization Bayes factor by a factor of $e^2$ or more. 

\begin{figure*}[tbh]

\includegraphics[width=0.68\columnwidth]{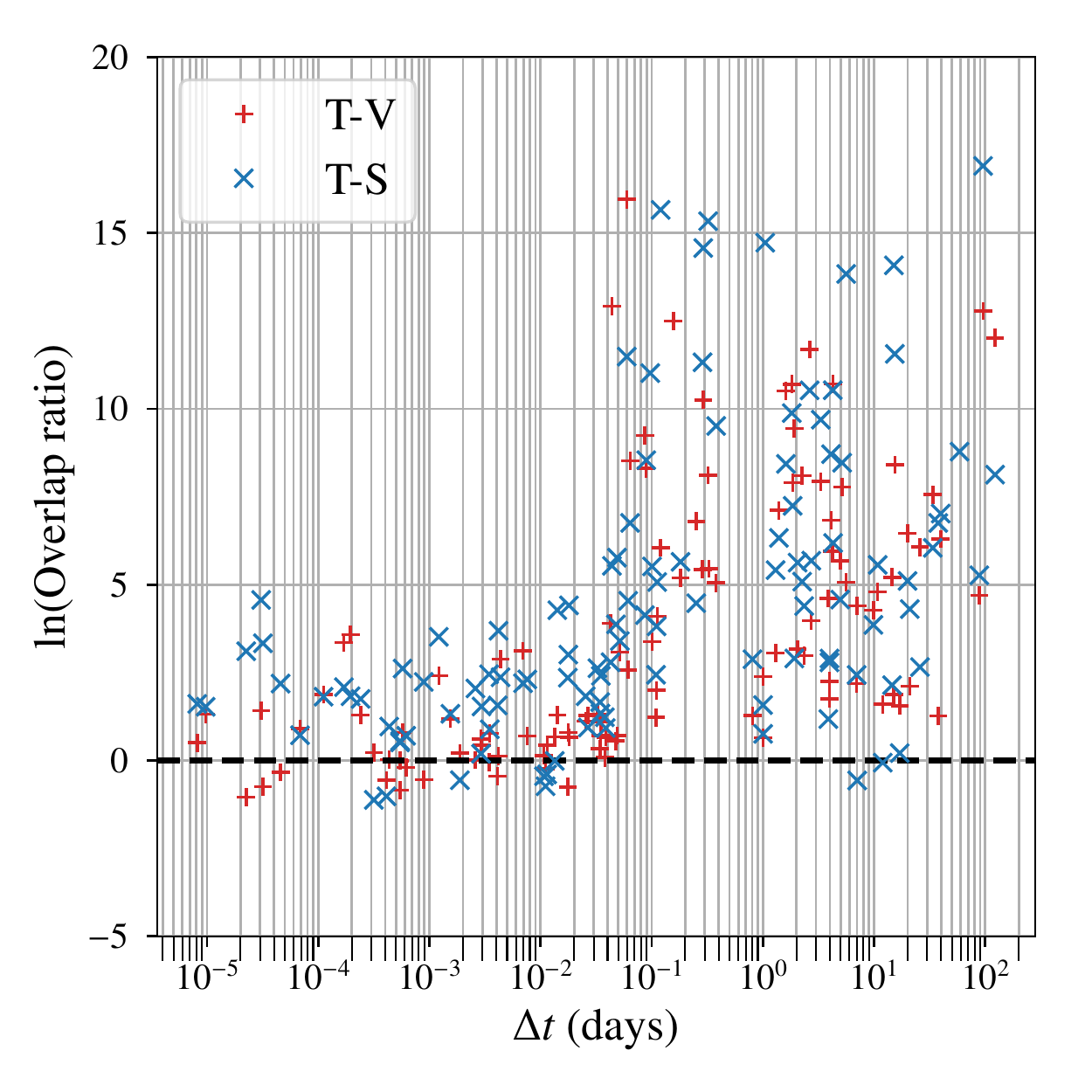}
\includegraphics[width=0.68\columnwidth]{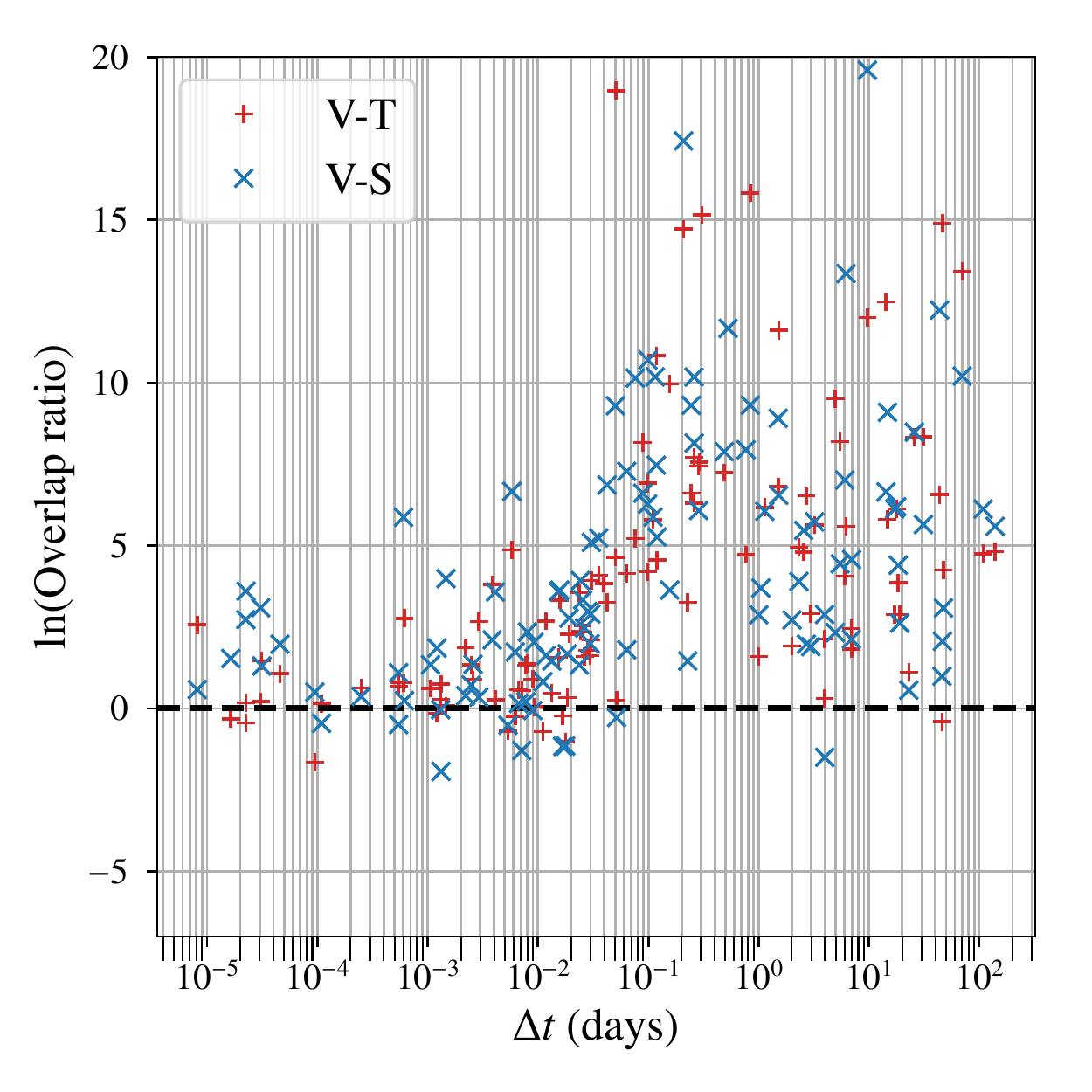}
\includegraphics[width=0.68\columnwidth]{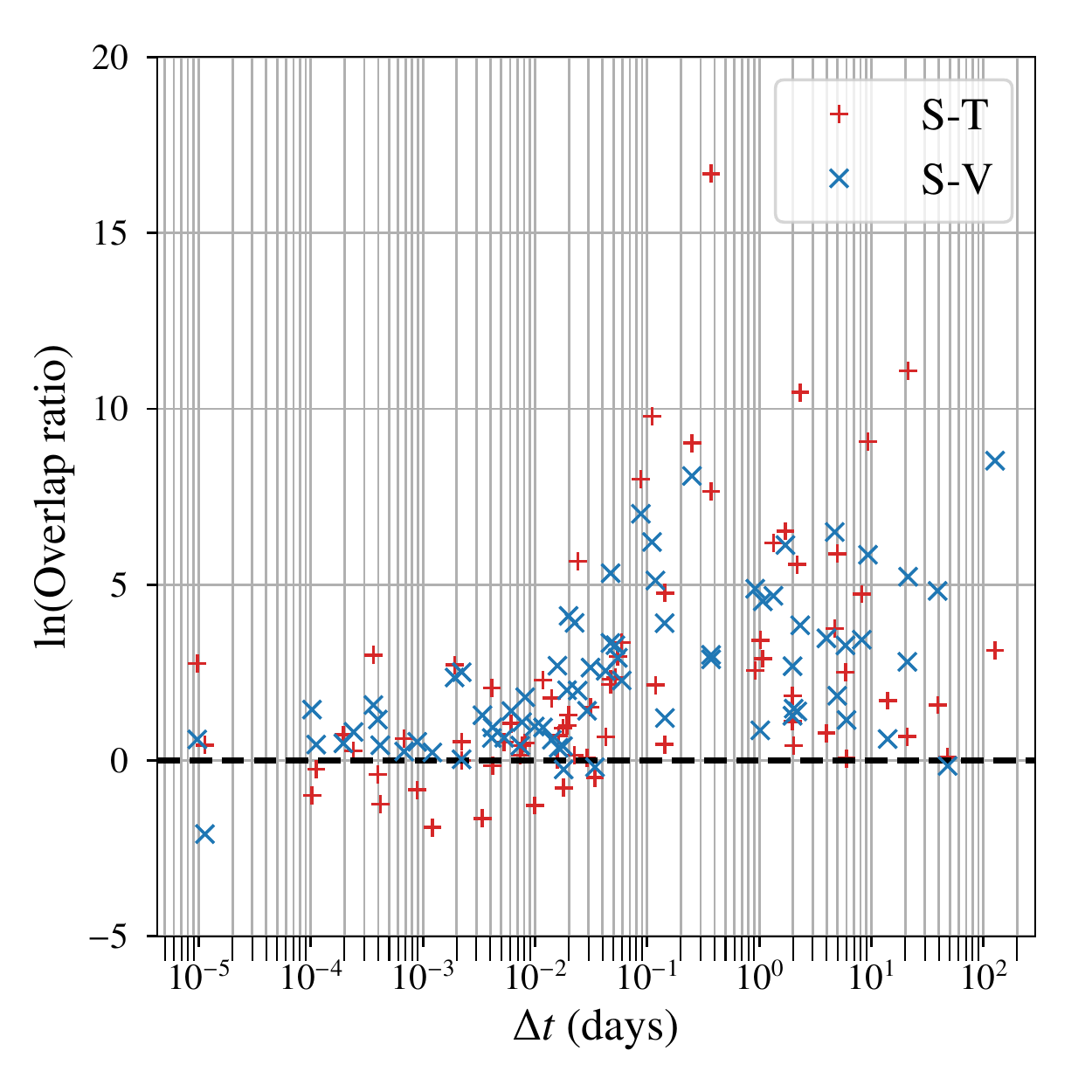}
\caption{ Correlation of the overlap ratios with the lensing time delay $\Delta t$ between the images for each set of injections, created using the tensor  (left), vector (middle) and scalar (right) polarization models. Different color markers show the overlap ratios between the ``right'' and ``wrong'' polarization models (for e.g., T-V denotes the overlap ratio between posteriors computed using the tensor and vector models). For the events below the black dashed lines, the posterior overlap ratio is less than one; hence lensing does not improve the polarization model selection.}
\label{fig:time_delay_corr}
\end{figure*}

The lower panel of Fig.~\ref{fig:overlap_ratio_dist} shows that, for a small fraction ($\sim 5-15\%$) of the simulated events, the posterior overlap ratio is less than (although very close to) one. That is, the lensing Bayes factor [Eq.~\eqref{eq:blu}] assuming the right polarization hypothesis is slightly smaller than the same assuming the wrong polarization hypothesis. These unusual event pairs do not show any significant correlations with the intrinsic or extrinsic parameters of the simulated BBHs.  However, these event pairs have small lensing time delays ($\sim$ less than half an hour). This is evident from Fig.~\ref{fig:time_delay_corr}, which plots  the log overlap ratios for all the lensed event pairs against the lensing time delay between these images.  This observation is broadly consistent with our expectation: during such short time delays ($\sim$ less than half an hour) the change in the antenna patterns of the detectors due to the rotation of the earth is negligible. Hence the antenna patterns at the times of the two images will be very similar to each other. Thus, the linear combination of the polarizations measured from the two events will be practically the same. In other words, the posteriors estimated from the two events will have high overlaps, irrespective of the polarization model used. For such events, lensing is not expected to bring significant additional improvements. This is clear from Fig.~\ref{fig:time_delay_corr}, which shows that the overlap ratios from the small-time-delay events are modest (less than $\sim e^4$). 

However, we would normally expect that the posterior overlaps of the right polarization model to be at least as large as the same using the wrong polarization models (in other words, the posterior overlap ratio should be greater than or equal to one). The reason for a small fraction of events to have overlap ratios slightly less than one is not well understood. It is likely that, when posteriors (using different polarizations) have very similar overlaps, the final results could be dominated by the numerical errors in our computations, such as the convergence of the parameter estimation and the inaccuracies in estimating the posterior distributions. We leave the detailed investigations on this as future work. 
  
\section{Summary and Future Work}
\label{sec:summary}
Probing the polarization content of the GWs observed by a network of ground-based detectors offers an interesting probe on the nature of gravity. While GR predicts only two (tensor) polarization modes, there are alternative theories that predict up to six polarizations (including scalar and vector modes, apart from the tensor modes predicted by GR). Each ground-based interferometer measures one particular linear combination of all these polarizations. Thus, if there are as many linearly independent detectors in the network as the number of independent polarizations, these polarizations modes can be extracted from the data, in principle. It turns out that the two scalar modes are degenerate as far as observations of ground-based detectors (which are quadrupolar antennas) are concerned. Thus five linearly independent polarization modes can be, in principle, extracted from the data of five linearly independent detectors. In practice, our ability to do this is limited by the presence of noise. In addition, the similar orientation of the two LIGO detectors in the USA makes this job difficult even with the upcoming network of five detectors including LIGO, Virgo, KAGRA and LIGO-India. 

Strong lensing of GWs can significantly improve our ability to constrain GW polarizations. Recent estimates suggest that $\sim 0.1-0.5\%$ of the GW signals from BBHs that the Advanced GW detectors will observe in the next few years will be strongly lensed by intervening galaxies, producing multiple ``images'' (copies) of the same signal that arrive at the detector with relative time delays of several minutes to weeks. Since several hundred BBH detections are expected in the next few years, the first observation of lensed GW signals is likely to happen soon. Since the wavelength of the GWs is significantly smaller than the mass scale of these lenses, lensing effects can be calculated using geometric optics. In this limit, lensing does not affect the frequency profile of the GW signals. Thus, the multiple images of a single merger will be comprised of the same GW polarizations (albeit with a relative magnification). Due to the rotation of the earth, each detected image will allow the GW detector network to measure different linear combinations of the same polarizations. This is effectively equivalent to multiplying the number of detectors to observe a single GW signal. 

We study the expected improvement, due to lensing, in our ability to probe the nature of GW polarizations making use of the Bayesian model selection formalism that was originally proposed by~\cite{Isi:2017fbj}. This uses a simplified model for the GW polarizations that are not present in GR: We assume that the time evolution of the additional polarizations (scalar and vector modes) follow that of the tensor modes. (Hence our ability to distinguish the polarization models depend greatly on the response of the GW detector network to different polarizations.) Additionally, we make a simplistic assumption that the GW polarizations consist of pure tensor, vector, or scalar modes. We show that strong lensing greatly improves our ability to distinguish the ``right'' and ``wrong'' polarization models for the GW signal. 

The joint Bayes factors  (likelihood ratio between two polarization models) for multiple, unrelated events can be obtained as the product of Bayes factors computed for individual events as the noise and the signal in the individual data segments  are unrelated. However, for a pair of strongly lensed events, though the noise is uncorrelated the GW signals present in the data segments are related. We show that the combined Bayes factor from such a lensed event is equal to the product of the individual Bayes factors and an additional factor, namely, \emph{posterior overlap ratio}, which is the ratio of the prior weighted overlaps of the posterior distributions of the GW parameters~\footnote{To be precise, the posterior distributions of the parameters that are expected to be common between the images.} that are computed assuming the two polarization models under consideration. From simulated BBH events in the three detector network consisting of Advanced LIGO and Virgo detectors in design sensitivity, we show that the overlap ratio for the majority of lensed events ($> 50\%$) is greater than {$e^2-e^3$}. This means that the Bayes factor supporting the right polarization hypothesis is improved by a factor of {$\sim 7-20$} for most of the lensed events (as compared to pairs of unlensed signals with similar strengths). The improvement can be as large as several thousands for about 10\% of the events. Note that, in this paper, we only consider lensing by galaxies, under the assumption that lensing probability of galaxy clusters is negligible. Lensing by clusters will introduce much larger time delays between images, thus significantly improving our ability to distinguish between polarization models (see, e.g., Fig.~\ref{fig:time_delay_corr}). 
The simplistic polarization models that we use in this paper can be extended to more realistic models, where the alternative model to GR would include scalar/vector modes \emph{in addition to} the tensor modes. Even if we assume that the scalar/vector modes follow the same phase evolution as the tensor modes, this will require us to model the effect of the binary's additional loss of energy and angular momentum (due to additional polarizations) on the orbital evolution itself. Additionally, the polarization model that is used in the model selection will require additional parameters that describe the relative strengths of the scalar, vector, and tensor modes (which will need to be marginalized away). Even then, the model selection described in Sec.~\ref{sec:model_sel_lens} can be used to characterize the expected improvement due to lensing. Since the ``right'' polarization model is expected to produce larger overlaps between the posteriors estimated from multiple lensed images, we expect that strong lensing will provide similar improvements in our ability to do model selection. Note that, in this paper we consider only double images produced by lensing,  while $\sim 30\%-40\%$ of the lensed events will also produce triple or quadruple images~\cite{Haris:2018vmn}, potentially providing further improvements in the polarization model selection. Such improvements in the effective number of detectors in the network might also enable us to perform polarization reconstruction in a model agnostic way. We plan to explore these aspects as follow-up projects.

\paragraph*{Acknowledgments:---} 
We are grateful to Max Isi, Bala Iyer, Shasvath Kapadia, Md Arif Shaikh, Apratim Ganguly, Kanhaiya Pandey, Aditya Vijaykumar, Soummyadip Basak and Mukesh Kumar Singh for useful discussions and to Otto Akseli Hannuksela for reviewing this manuscript. Our research was supported by the Department of Atomic Energy, Government of India. PA's research was funded by the Max Planck Society through a Max Planck Partner Group at ICTS-TIFR and by the Canadian Institute for Advanced Research through the CIFAR Azrieli Global Scholars program. The numerical calculations reported in the paper are performed on the Alice computing cluster at ICTS-TIFR, with the aid of \textsc{LALSuite} and \textsc{PyCBC} software packages.

\bibliography{LensingPol}
\end{document}